# Influence of doping on the properties of vanadium oxide gel films

A. L. Pergament, A. A. Velichko, O. Ya. Berezina, E. L. Kazakova, N. A. Kuldin and D. V. Artyukhin

Physics and Technology Department, Petrozavodsk State University,
Petrozavodsk, 185910, Russian Federation

e-mail: aperg@psu.karelia.ru

**Abstract.** Effect of doping with H and W on the properties of $V_2O_5$ and $VO_2$ derived from $V_2O_5$ gel has been studied. It is shown that the treatment of $V_2O_5$ in low-temperature RF hydrogen plasma for 1 to 10 min. leads to either hydration of vanadium pentoxide or its reduction (depending on the treatment conditions) to lower vanadium oxides. For some samples, which are subject to plasma treatment in the discharge active zone, a non-ordinary temperature dependence of resistance, with a maximum at $T \sim 100$ K, is observed. For W-doped $VO_2$ films, it is shown that substitution of $V^{4+}$ with $W^{6+}$ results in a decrease of the temperature of metal-insulator transition. Also, it has been shown that the doping of the initial films with ~3 at.% of W reduces the statistical scatter in the threshold parameters of the switching devices with S-shaped *I-V* characteristics on the basis of $V_2O_5$ gel films.

Vanadium oxides are of interest due to metal-insulator transitions (MIT) occurring in many of them at different temperatures [1]. Doping with various elements allows a controllable change of the properties of these materials and the parameters of their phase transitions (transition temperature $T_t$, hysteresis loop width, etc). These investigations are of importance both for better understanding the MIT mechanism and for potential applications, since the transition-metal oxides are the very materials which seem to underlie of oxide electronics [2]. In the present work we report on modification of electrical properties of hydrated vanadium pentoxide under the action of hydrogen RF plasma treatment, vacuum annealing and doping with tungsten.

Initial film samples were prepared by the sol-gel method. This method is currently considered as a novel and efficient technique to prepare thin films of diverse materials, first of all – oxides; in this process, thin films of metal oxides can be deposited directly on the immersed substrate or by spin coating [3-6]. Among other advantages of this method one can point out that it is easy to apply to various kinds of substrates with a large surface area and complex surface shape [3]. Also, an essential merit of the sol-gel method is the easiness of doping – merely by adding corresponding quantities of a certain composition into the prepared sol.

Vanadium pentoxide gel solution was prepared by the quenching method [3]. Tungsten doped gels were obtained by adding of $WO_3$ powder directly into the vanadium pentoxide melt [4]. It had been found experimentally that the dissolution limit of $WO_3$ in $V_2O_5$ melt corresponded to 12 at.% of tungsten. In order to deposit a film, a few millilitres of the gel solution were placed onto a substrate and excess solvent was allowed to evaporate at room temperature for 24 hours. As a result, yellowish-brown xerogel films were prepared with a typical thickness $d \sim 1$ to 5 μm. These dried in air films represented X-ray amorphous $V_2O_5 \times nH_2O$ with a layered structure [5].

X-ray diffraction pattern of the initial $V_2O_5$-gel film is presented in Fig.1; the interlayer spacing is 11.5 Å (measured from the (001) diffraction peak at $s = 0.55$ Å$^{-1}$), which corresponds to the water content $n=1.6–1.8$ [3,5]. Water can be removed upon heating to $n=0.1 \div 0$ at $T = 210$-270ºC. Amorphous $V_2O_5$ is obtained at this temperature, and crystallization into orthorhombic $V_2O_5$ occurs around 350°C [3]. Vanadium dioxide can be easily prepared by a thermal treatment at $T \sim 500°C$ in vacuum or a reducing atmosphere [4-6]. The formation of $VO_2$ rather than other lower vanadium oxides is accounted for by its thermodynamic properties [7], and preparation of other vanadium oxides in a thin film form is very difficult due to the narrowness of the stability range of any oxide [3,7].

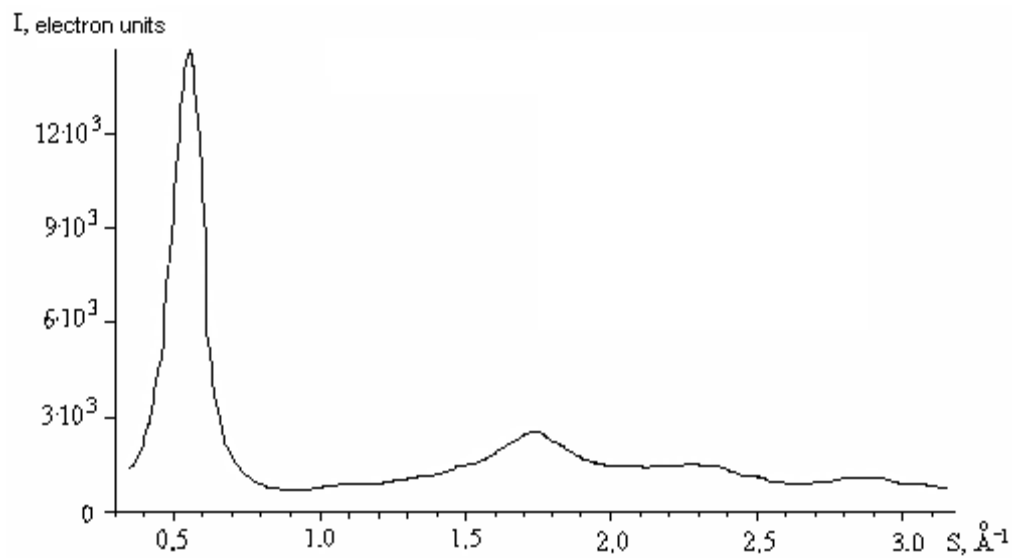

**Figure 1**. X-ray intensity as a function of diffraction vector $s = 4\pi\sin\theta/\lambda$ (where $\theta$ is the diffraction angle, and $\lambda=0.7107$ Å – Mo-$K_\alpha$ radiation) for $V_2O_5 \times nH_2O$ film.

The possibility of doping with hydrogen has been studied with the samples of amorphous $V_2O_5$, obtained by the thermal treatment of an initial $V_2O_5 \times nH_2O$ film in air at $T = 250°C$. Plasma treatment has been performed in an RF (2.45 GHz) hydrogen reactor with remote plasma [8]. Design features of the reactor exclude the sample heating, permitting however its contact with low-energy $H^+$ ions and atomic hydrogen.

First, we present the experimental results obtained at various pressure $p$ (10 – 100 Pa) in the reactor, time $t$ of treatment (20 sec. to 10 min.), and the distance $l$ of the treated sample from the discharge active zone (0 – 10 cm). At $t < 1$ min. and $l = 5 – 10$ cm, an effective hydrogen insertion has been observed: the appearance of internal electrochromic effect [3] reveals the hydrogen intercalation into $V_2O_5$. This internal electrochromic effect is associated with the hydrogen redis-tribution inside the $V_2O_5$-gel film; that is, the process of plasma treatment results in formation of polyvanadic acid $H_xV_2O_5 \times nH_2O$ with x = 0.3-0.4 [3]. It is pertinent to mention here that the insertion of hydrogen into vanadium oxides with formation of vanadium oxide bronzes is not an emergent phenomenon: it is observed during electro- and photochromism [9,10], and even merely by means of immersion of a vanadium oxide film into a hydrogen-containing liquid [11].

Apropos of RF plasma treatment, we encountered the fact that varying the parameters ($p$, $t$, $l$) it was possible to obtain any lower vanadium oxide. As an example, the resistance temperature dependence for one of the samples, presented in Fig.2 (curve 1), shows the MIT at $T_t \approx 170$ K which is characteristic of the $V_6O_{11}$ phase of the $V_nO_{2n-1}$ series [1].

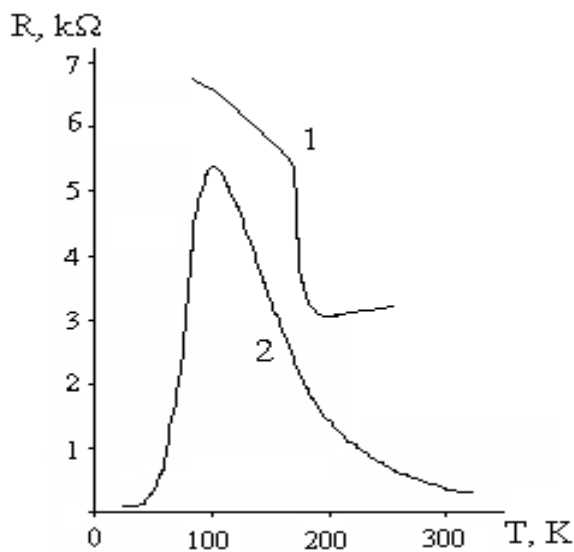

**Figure 2.** Temperature dependence of resistance (on cooling) of the vanadium oxide samples subject to plasma treatment (1) for $t$ = 1 min. in the centre of waveguide ($l$ = 5 cm), and (2) for $t$ = 2 min. directly in the discharge ($l$ = 0).

Moreover, what happens on placing the sample directly into the discharge is remarkable. Curve 2 in Fig.2 demonstrates the temperature dependence of resistance for the sample, which is plasma-treated in the centre of discharge ($l \approx 0$) at $p$ = 100 Pa. This result is rather unforeseen: such an $R(T)$ curve with a maximum has never been reported for vanadium oxides heretofore. In the high temperature region this dependence has a semiconducting character followed by the sharp resistance fall at $T < T_m \approx 100$ K by almost two orders of magnitude, and the specific resistivity at $T = 20$ K has been estimated to be $\rho \approx 0.1$ $\Omega \cdot$cm for the film thickness 5 μm.

The above-described behaviour of the *R(T)* dependence might be attributed to a re-entrant (or "inverse") MIT. The second high-temperature transition in $(V_{1-x}Cr_x)_2O_3$ (besides the first "usual" MIT at $T_t = 150$ K) [1,12,13] is an example of such an inverse MIT: the metallic phase is low-temperature, and the high-temperature phase is insulating. Another example of inverse MIT is a more spread-out transformation in $NiS_{2-x}Se_x$, for which $T_t$ ranges from 30 to 100 K with variation of x. Also, $BaCo_{0.9}Ni_{0.1}S_{2-y}$, europium oxide, CMR manganites, and some other materials demonstrate that same behaviour – see the pertinent surveys [1,12-14] and references therein. However, the experimental curve (#2 in Fig.2) might also be an indication of superconductivity (SC): a similar dependence of *R* on *T* is characteristic of some under- and over-doped HTSC cuprates (such as, e.g., $TlCa_{1-x}Nd_xSr_2CuO_4$ at x=0.75 [12]).

As regards to the phase composition of those samples, treated in low-temperature RF hydrogen plasma, X-ray analysis shows a mixture of lower vanadium oxides, which are difficult to distinguish. Presumably, if we do deal with a vanadium-based SC phase [15,16], it might be a certain hydrogen bronze $H_xV_nO_{2n-1}$ of a Magneli-phase lower vanadium oxide.

Next, we have doped the vanadium oxide gel with tungsten as described above (see also ref.[4] where the process of doping is described in more detail). The films deposited from such a gel represent the following composition: $V_{2-y}W_yO_{5\pm\delta} \times nH_2O$, where y is varied from zero (pure $V_2O_5$-gel) to 0.12, and *n* = 1.6–1.8. These films, after vacuum annealing, transform into vanadium dioxide doped with tungsten. Influence of doping on the electrical properties of $VO_2$ is demonstrated by Fig.3. An increase in the W content leads to a decrease in the resistivity of the sample and in a shift of the transition temperature $T_t$ toward the low-temperature region. In addition, some modification of the hysteresis loop parameters is observed, as it is discussed in the work [4]. At the impurity content more than 6 at.%, the MIT no longer occurs, though the *R*(*T*) dependence remains still semiconducting (Fig.3, curve 3). For complete metallization of $VO_2$, the concentration of W dopant should be ~14 at.% [17].

It is understandable that "metallic" $VO_2$ (i.e. $V_{1-x}W_xO_2$ with x=0.14 [17] or $H_xVO_2$ with x=0.04 [11]) could serve as a merely precursor to search for superconductivity [15,16]. Experience shows that all the known HTSC materials are strongly correlated metals (in normal state) and that they are prone to MIT. The next stage should be to intercalate some additional elements (e.g. hydrogen, alkali or alkali-earth atoms) into this host precursor structure. Light electronegative elements promote an enhancement of the ionicity bond degree [18] and contribute to the softening of the phonon modes, which is important for the appearance of HTSC. The scepticism of scientific community with regard to the search for new SC materials amid non-Cu-based (particularly, V-based) compounds[1] is founded, among others arguments, upon the opinion that "binary structures with only two sites per unit cells luck the possibility of being able to substitute and introduce charge carriers without introducing disorder directly into the conduction band, and thus are considered to be less likely candidates for supporting SC than more complex structures" [15]. However, the ability of vanadium oxides to be substituted (e.g., V → W or Mo) and intercalated (by alkali and alkaline-earth atoms) simultaneously [20,21] seems to remove this argument. In addition, according to the authors of [17], the samples of $V_{1-x}W_xO_2$ with x up to 0.14 remain quite ordered with undistorted rutile structure.

Finally, we have studied the switching effect in the MDM structures on the basis of W-doped $V_2O_5$-gel films. These structures require preliminary electroforming [3] resulting in formation of a conducting channel within the initial dielectric film. The channel consists of vanadium dioxide and switching is associated with the insulator-to-metal transition in $VO_2$. As was shown in

---

[1] Note that a copper-less HTSC material La-O(F)-Fe-As has been discovered recently [19].

[3], for the pure (undoped) $V_2O_5$-gel, the switching parameters vary in a wide range, because the phenomenon of electrical forming is statistical in nature (likewise electrical breakdown), and its statistical character shows itself as a spread in the observed threshold voltages of the ensuing structures.

In this work, we have found that the statistical spread in values of the threshold parameters can be minimized by means of doping (Fig.4). Unsophisticated reasoning could account for this result: For pure $V_2O_5$-gel, electroforming leads to formation of $VO_2$ with $T_t = 340$ K, and each switching event requires heating up and cooling down from room temperature to $T_t$; this thermal cycling can modify the peripheral regions of the channel, which leads to the change of the switching threshold voltage. On the other hand, for the samples with a high concentration of W, the MIT is suppressed (see Fig. 3, curve 3), and switching is degenerated. Between these two outermost points, an optimum should exist. As one can see from Fig.4, this optimal value corresponds to 3 at.% of W, i.e. for $V_{0.97}W_{0.03}O_2$ the switching parameters are the most stable. The transition temperature of $V_{0.97}W_{0.03}O_2$ (Fig.3, curve 2) is lower than that for pure vanadium dioxide, and hence the magnitude of the aforesaid thermal cycling is decreased, which results in an increase of stability of the switching parameters.

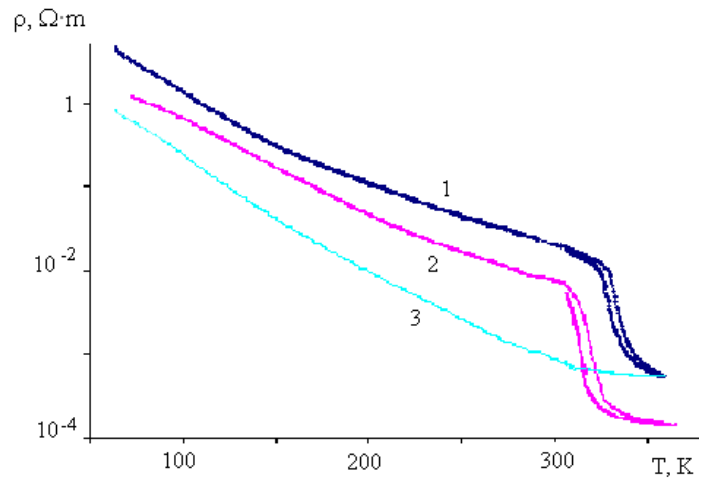

**Figure 3**. Resitivity *vs*. temperature for vanadium dioxide (1) and for $VO_2$ doped with 3 at.% (2) and 12 at.% (3) of tungsten.

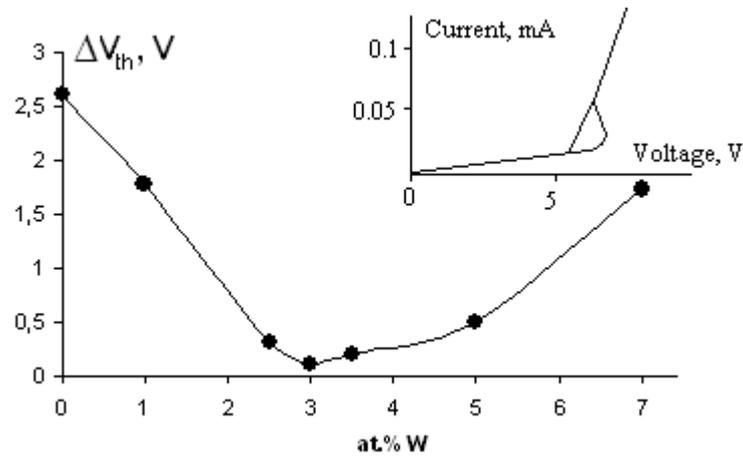

**Figure 4.** Threshold range ($V_{th}$ scattering) as a function of W admixture percentage. A typical *I-V* curve of Au–$V_{2-y}W_yO_5 \times nH_2O$–Au sandwich structure with y=0.03 is shown in the insert. Thickness for all the samples is *d* ~ 1 μm and the mean value of $V_{th}$ is actually independent of W concentration ($<V_{th}>$= 7.1±1.5 V with reliability 0.95).

To summarize, it is shown that doping with tungsten and hydrogen affects severely on the electrical properties of vanadium oxides. The treatment of $V_2O_5$ in low-temperature hydrogen plasma leads to either hydration of vanadium pentoxide or its reduction to lower vanadium oxides. Doping of $VO_2$ with W leads to a decrease of the transition temperature, which is in accordance with the literature data [4,17] and supports the Mott mechanism of MIT in vanadium dioxide [13,22]. The introduction of ~3 at.% of W reduces the scatter in the threshold voltage of the switching MDM structures on the basis of $V_2O_5$ gel.

Acknowledgments. This work was supported by the Svenska Institutet (Dnr: 01370/2006), the Federal Agency for Science and Innovations of R.F. (contract 02.513.11.3351), the Ministry of Education of R.F. and the U.S. Civilian Research and Development Foundation (CRDF award No.Y5-P-13-01).